# Effect of 3d Transition Metal Doping on the superconductivity in Quaternary Fluoroarsenide CaFeAsF

Satoru Matsuishi[1], Yasunori Inoue[2], Takatoshi Nomura[2], Youichi Kamihara[3], Masahiro Hirano[1,2] and Hideo Hosono[1,2,4]

[1]Frontier Research Center, Tokyo Institute of Technology, 4259 Nagatsuta-cho, Midori-ku, Yokohama, Japan

[2]Materials and Structure Laboratory, Tokyo Institute of Technology, 4259 Nagatsuta-cho, Midori-ku, Yokohama, Japan

[3]TRIP, JST in Materials and Structure Laboratory, Tokyo Institute of Technology, 4259 Nagatsuta-cho, Midori-ku, Yokohama, Japan

[4]ERATO-SORST, JST in Frontier Research Center, Tokyo Institute of Technology, 4259 Nagatsuta-cho, Midori-ku, Yokohama, Japan

E-mail: :satoru@lucid.msl.titech.ac.jp

**Abstract** We examined doping effect of 3d transition metal elements (*TM*: Cr, Mn, Co, Ni, and Cu) at the Fe site of a quaternary fluoroarsenide CaFeAsF, an analogue of 1111-type parent compound LaFeAsO. The anomaly at ~120 K observed in resistivity ($\rho$) vs. temperature ($T$) plot for the parent compound is suppressed by the doping of each *TM* element. Furthermore, Co- and Ni-doping (CaFe$_{1-x}$*TM*$_x$AsF, *TM* = Co, Ni) induces superconductivity with a transition temperature maximized at the nominal $x$ = 0.10 for Co (22 K) and at $x$ = 0.05 for Ni (12 K). These optimal doping levels may be understood by considering that Ni$^{2+}$(3d$^8$) adds double electrons to the FeAs layers compared with Co$^{2+}$ (3d$^7$). Increased $x$ for Co or Ni breaks the superconductivity while metallic nature $d\rho/dT > 0$ is still kept. These observations indicate that Co and Ni work as electron donors. In contrast, neither of Cr, Mn nor Cu-doping induce superconductivity, yielding $d\rho/dT < 0$ in the below the $\rho$-$T$ anomaly temperature, indicating that these transition metal ions act as scattering centers. The two different behavior of TM replacing the Fe site is discussed in relation to the changes in the lattice constants with the doping.

Effect of 3d Transition Metal Doping on the Superconductivity in CaFeAsF

Effect of 3d Transition Metal Doping on the Superconductivity in CaFeAsF

**1. Introduction**

The discovery of superconductivity in fluorine-doped LaFeAsO with transition temperature $T_c$ = 26 K [1] triggered intensive studies of FeAs-based and related layered compound systems including $R$FeAsO ($R$ = rare-earth) [2]-[8], $B$Fe$_2$As$_2$ ($B$ = alkali-earth) [9]-[11], $A$FeAs ($A$ = alkali) [12] and Fe$Ch$ ($Ch$ = chalcogen) [13]-[15] with a hope for realizing higher-$T_c$ superconductors. These efforts lead to raising $T_c$ up to 56 K in Th-doped GdFeAsO so far [8]. These superconductors belong to ZrCuSiAs-type (space group P4/nmm), ThCr$_2$Si$_2$-type (I4/mmm), Cu$_2$Sb-type (P4/nmm) or PbO-type (P4/mmm) consisting of an alternating stack of a basal Fe layer (FeAs)$^{\delta-}$ and a blocking layer ($RO$)$^{\delta+}$, $B^{\delta+}$ or $A^{\delta+}$ layers except the Fe$Ch$ without blocking layer. They suffer a crystallographic transition from the tetragonal to orthogonal, accompanying with antiferromagnetic transition yielding anomaly in electrical resistivity at 140-200 K [1],[3],[7]-[11],[16]-[17]. It is now a consensus that the superconductivity in all the compounds is induced as a result of electron or hole doping to the (FeAs)$^{\delta-}$ layer, which simultaneously suppress both transitions to occur. Thus, exploring efforts in the material studies have been focused on the synthesis of ZrCuSiAs- and related-type compounds containing the square iron lattice as well as on the carrier doping technique [19].

Four types doping methods have been reported to date i.e., (1) indirect electron-doping by an introduction of fluorine into the O sites and the oxygen-vacancy formation in insulating layers ($RO$)$^{\delta+}$ in $R$FeAsO [20], [21], (2) indirect hole-doping by alkali-metal doping to $B^{\delta+}$ layer in $B$Fe$_2$As$_2$ [9]-[11] and formation of vacancy at $A$ site in $A$FeAs [12], (3) electron-doping by partial replacement of Fe with Co in $R$FeAsO and $A$Fe$_2$As$_2$ [22]-[26] and (4) direct electron-doping by formation of vacancy at the Se or Te sites in Fe$Ch$.[13]-[[15]. The direct hole-doping has not been achieved yet.

Recently, we reported synthesis of a new FeAs containing ZrCuSiAs-type compounds, $A$FeAsF ($A$ = Ca, Sr), in which the (FeAs)$^{\delta-}$ layer is sandwiched by ($AF$)$^{\delta+}$ layer in place of ($RO$)$^{\delta+}$ layer in $R$FeAsO compounds (figure 1a) and their superconductivity induced by partial substitution of Fe with cobalt [27],[28]. Following these report, the existence of $A$FeAsF ($A$ = Ca, Sr and Eu) system was reported by a few groups [29], [30] and superconductivity induced by lanthanide metal doping with $T_c$ up to 56 K have been posted on arXiv/condmat preprint server [31],[32]. For the case of partial substitution of

Effect of 3d Transition Metal Doping on the Superconductivity in CaFeAsF

Fe with Co, the optimal $T_c$ is 22 K for 10 % Co-substituted CaFeAsF (CaFe$_{0.9}$Co$_{0.1}$AsF) and 4 K for SrFe$_{0.875}$Co$_{0.125}$AsF. $T_c$ in the former is comparable to those in Co-substituted FeAs-based compounds. ($T_c$ = 14 K for LaFeAsO, 17 K for SmFeAsO, 20 K for SrFe$_2$As$_2$, and 22 K for BaFe$_2$As$_2$) [22]-[26]. Our discovery indicates that the Co-doping technique is a universal way to convert the FeAs-based layered compounds to superconductors. Further, the effectiveness of Co-doping suggests that the square Fe lattice in the FeAs layer is much more robust to impurities, than CuO$_4$ planes in High-$T_c$ cuprate. [33]

In this study, we examined substitution of a series of 3d-transtion metals to the iron site on the emergence of superconductivity in CaFeAsF. Temperature dependence of electrical resistivity and lattice parameter changes upon substitution were measured. As a consequence, it was revealed that Ni-substitution induces superconductivity similar to the Co while Cr-, Mn- or Cu-substitution does not yield superconductivity. The optimal concentration for Ni was almost a half of Co, suggesting that both ions with excess 3d electron serve as electron donor.

## 2. Experimental

Samples were prepared by a solid state reaction of CaF$_2$ (99.99 %), CaAs, Fe$_2$As and $M_2$As ($M$ = Cr, Mn, Co, Ni, or Cu): CaF$_2$ + CaAs + (1−$x$) Fe$_2$As + $x$ $M_2$As → 2CaFe$_{1-x}M_x$AsF. CaAs was synthesized by heating a mixture of Ca shots (99.99 wt. %) and As powders (99.9999 wt. %) at 650°C for 10 h in an evacuated silica tube. Fe$_2$As and $M_2$As were synthesized from powders of respective elements at 800 °C for 10 h (Fe: 99.9 wt. %; Cr: 99 wt. %, Mn: 99.99 wt. %, Co: 99.9 wt. %, Ni: 99.99 wt. %, Cu: 99.9 wt.%). These products were then mixed in stoichiometric ratios, pressed, and heated in evacuated silica tubes at 1000 °C for 10 h to obtain sintered pellets. All the procedures until the sealing to silica glass tubes were carried out in an Ar-filled glove box (O$_2$, H$_2$O < 1 ppm).

The crystal structure and lattice constants of the materials were examined by powder X-ray diffraction (XRD; Bruker D8 Advance TXS) using Cu K$\alpha$ radiation from a rotating anode with an aid of Rietveld refinement using Code TOPAS3 [34]. Temperature dependence of DC electrical resistivity ($\rho$) at 2-300 K was measured by a four-probe technique using platinum electrodes deposited on samples

## 3. Results and Discussion

Figure 1b shows powder XRD pattern of non-doped CaFeAsF. Except the several weak peaks arising from impurity phase ($Fe_2As$, the volume fraction being 2 % at most), each of the major peaks was assigned to the CaFeAsF phase and room temperature lattice constants was evaluated as $a$ = 0.3879 nm and $c$ = 0.8593 nm. Figure 1c shows temperature dependence of electrical resistivity of CaFeAsF. With a decrease in temperature, ρ-$T$ curves exhibit sudden decreases at ~120 K ($T_{anorm}$). This anomalous behavior is quite analogous to those in $R$FeAsO and $B$Fe$_2$As$_2$, implying that the crystallographic transition takes place at $T_{anorm}$. CaFeAsF also likely suffers a magnetic ordering in the similar temperature region.

Figure 2 and 3 show powder XRD patterns of CaFeAsF doped with Cr (figure 2a), Mn (2b), Co (3a), Ni (3b), and Cu (3c). Except the small peaks attributable to those of impurity phases, most of the pesks in all the patterns are assigned to originate from CaFeAsF phase (volume fraction: > 97% for Cr-doping, > 79% for Mn, > 83% for Co, > 88 % for Ni, and 80% for Cu,). FeAs (< 10%) and $CaF_2$ (< 12%) were mainly observed as impurity phases and their volume fractions were not depend on transition-metal substitution level. It indicates that these impurity phases may result from the loss of Ca elements due to vaporization from the starting mixture during heating process. For cases of Mn-, Co-, and Ni-doping, oxide impurities CaO and $Fe_2O_3$ (volume fraction: < 7%) were observed, indicating oxygen contamination of calcium or transition metal reagents during heating process. Mn-substitution over 15 % leads to the segregation of $CaMn_2As_2$ phase, indicating the Mn concentration exceeds the solubility limit. Since all the samples contain impurity phases, practical doping level $x$ is not obvious. Therefore, we used the molar fractions of transition metals in starting mixture as nominal $x$ hereafter.

Figure 4 shows $a$- and $c$-axis lengths of CaFe$_{1-x}$M$_x$AsF ($M$ = Cr, Mn, Co, Ni and Cu) as functions of nominal $x$. The $a$-axis length evidently increases with Mn- and Cu-substitution (+0.10% for Mn and +0.23% for Cu-doping for nominal $x$ = 0.05), while it decrease with Cr-substitution (-0.09% for nominal $x$ = 0.05). Since the $a$-axis length directly corresponds to the distance between first neighboring irons in the iron square

Effect of 3d Transition Metal Doping on the Superconductivity in CaFeAsF

lattice ($r_{\text{Fe-Fe}} = a/\sqrt{2}$, See figure 1a), these result indicate that Cr, Mn and Cu-substitution introduce the lattice distortion yielding the change in the Fe-Fe distance. In contrast, the increment of *a*-axis length due to Co- and Ni-substitution is smaller than that for the above cases (+0.04% for Co and +0.03% for Ni-doping for nominal $x$ = 0.05), indicating Co- and Ni-substitution don't induce large distortion in the iron lattice. As shown in lower column of figure 4, the *c*-axis length increases with Cr- and Mn-doping and decreases with Co- and Ni-doping. This may be understood as a consequence of the reduction or enhancement in Coulombic interaction between the $(\text{CaF})^{\delta+}$ and $(\text{FeAs})^{\delta-}$ layers upon doping, providing the evidence that the Cr/Mn and Co/Ni substitution respectively decrease and increase the effective electron population i.e, hole /electron-doping in the FeAs layers.

Figure 5 shows ρ-*T* curves for Cr- and Mn-doped samples. With an increase of nominal $x$ for both Cr and Mn, the ρ-*T* anomaly shifts to lower temperatures (see a shoulder below 60 K for nominal $x$ = 0.05 for Cr and 0.10 for Mn) and eventually suppressed for nominal $x$ > 0.14 for Cr and > 0.1 for Mn. However, the Cr- and Mn-substitution enhance the resistivity and change the temperature coefficient at low temperature from positive to negative. These observations indicate that Cr and Mn form scattering center, possible magnetic, disturbing the electron conduction in the FeAs layer. Especially, the Mn-doping induces the higher resistivity, presumably corresponding to larger structural changes in the FeAs-layer.

Figure 6 shows temperature dependences of resistivity for $\text{CaFe}_{1-x}\text{Co}_x\text{AsF}$ and $\text{CaFe}_{1-x}\text{Ni}_x\text{AsF}$. It is clearly demonstrated that the ρ-*T* anomaly temperature is lowered with an increase in Co and Ni contents and superconductivity is induced for nominal $x$ > 0.07 for the Co-doping and nominal $x$ > 0.05 for the Ni-doping. On the contrary, the Cu-doping never induces superconductivity and the electrical resistivity at low temperatures is enhanced by ~$10^3$ times than that of non-doped sample. Cu apparently acts as a strong scattering center for the itinerant electrons.

Figure 7a and 7b shows close-up views of ρ-*T* curves around onset superconducting transition temperatures ($T_{\text{onset}}$) for Co and Ni-substituted CaFeAsF samples. $T_{\text{onset}}$ as a functions of nominal $x$, obtained from the ρ-*T* plots are summarized in figure 7c. It is evident that the optimal Ni-doping level yielding maximum $T_{\text{onset}}$ (nominal $x$ ~ 0.05) is nearly half of that for the Co-doping (nominal $x \sim 0.1$). That suggests that $Co^{2+}$ with $3d^7$ electronic configuration gives an additional electron, while $Ni^{2+}$ with $3d^8$ gives two electrons to FeAs-layer. It is noteworthy that Co acts as a non-magnetic donor for small values of nominal $x$ although LaCoAsO exhibit ferromagnetism [35]. Further, it is of interest to note that the threshold and optimal electron-doping level (~ 0.1 electron / Fe) is close to that in $R$FeAs($O_{1-x}F_x$) system notwithstanding that the impurity doped layer is different.

Finally, we would like to stress the practical importance of Co/Ni-doping in fabrication of epitaxial thin films of FeAs-based superconductors. Carrier doping is requisite for emergence of superconductivity in this system. Several methods have been reported to be effective to date such as electron-doping via the formation of oxygen vacancy for the 1111 system and hole-doing via substitution of alkaline earth ion sites with an alkali ion for the 122 system. However, neither of them was practically hard to be doped to the thin films by vapor phase deposition processes due presumably to weak chemical bonding strength of these species with the host lattice. Using Co as dopant, the first demonstration of epitaxial thin films of $Sr(Fe,Co)_2As_2$ exhibiting a $T_c = 22$ K have been realized recently [36].

## 4. Summary

We examined the partial replacement of Fe site in CaFeAsF with 3d-transition metals (TM: Cr, Mn, Co, Ni and Cu) and obtained the following conclusion;

(1) Only Co- or Ni-doping was effective for emergence of superconductivity. The optimal doping level for Ni was close to a half of that for Co-doping.

(2) Cr-, Mn- or Cu-doping led to enhanced resistivity at low temperatures, without inducing superconductivity, indicating that these ions act as scattering centers.

(3) Spacing between CaF and FeAs layers changed with TM-doping and the sign of the change was reverse between Co/Ni and Cr/Mn groups. This result was understood by different polarity in carriers (hole or electron) doped to the FeAs layer.

## References

[1] Kamihara Y, Watanabe T, Hirano M, and Hosono H 2008 *J. Am. Chem. Soc.* **130** 3296-3297

[2] Takahashi H, Igawa K, Arii K, Kamihara Y, Hirano M and Hosono H 2008 *Nature* **453** 376-378

[3] Chen G F, Li Z, Wu D, Li G, Hu W Z, Dong J, Zheng P, Luo J L and Wang N L 2008 *Phys. Rev. Lett.* **100** 247002

[4] Ren Z-A, Yang J, Lu W, Yi W, Che G-C, Dong X-L, Sun L-L and Zhao Z-X 2008 *Mater. Res. Innov.* **12** 105

[5] Ren Z-A, Yang J, Lu W, Yi W, Shen X-L, Li Z-C, Che G-C, Dong X-L, Sun L-L, Zhou F and Zhao Z-X 2008 *Europhys. Lett.* **82** 57002

[6] Chen X H, Wu T, Wu G, Liu R H, Chen H and Fang D F 2008 *Nature* **453** 761

[7] Ren Z-A, Lu W, Yang J, Yi W, Shen X-L, Li Z-C, Che G-C, Dong X-L, Sun L-L, Zhou F and Zhao Z-X 2008 *Chin. Phys. Lett.* **25** 2215

[8] Wang C, Li L, Chi S, Zhu Z, Ren Z, Li Y, Wang Y, Lin X, Luo Y, Jiang S, Xu X, Cao G and Xu Z 2008 *Europhy. Lett.* **83** 67006

[9] Rotter M, Tegel M and Johrendt D 2008 *Phys. Rev. Lett.* **101** 107006

[10] Chen G-F, Li Z, Li G, Hu W-Z, Dong J, Zhou J, Zhang X-D, Zheng P, Wang N-L and Luo J-L 2008 *Chin. Phys. Lett.* **25** 3403

[11] Wu G, Chen H, Wu T, Xie Y-L, Yan Y-J, Liu R-H, Wang X-F, JYing J-J and Chen X-H 2008 *J. Phy.: Cond. Matter* **20** 422201

[12] Wang X C, Liu Q Q, Lv Y X, Gao W B, Yang L X, Yu R C, Li F Y and Jin C Q 2008 *Solid State Communication* **148** 538

[13] Hsu F-C, Luo J-Y, Yeh K-W, Chen T-K, Huang T-W, Wu P M, Lee Y-C, Huang Y-L, Chu Y-Y, Yan D-C, Wu M-K 2008 *Proc. Natl. Acad. Sci. USA* **105** 14262

[14] Mizuguchi Y, Tomioka F, Tsuda S, Yamaguchi T and Takano Y 2008 *Appl. Phys. Lett.* **93**, 152505

[15] Yeh K-W, Huang T-W, Huang Y-L, Chen T-K, Hsu F-C, Wu P M, Lee Y-C, Chu Y-Y, Chen C-L, Luo J-Y, Yan D-C and Wu M-K 2008 *Europhys. Lett.* **84** 37002

[16] de la Cruz C, Huang Q, Lynn J W, Li J, Ratcliff II W, Zarestky J L, Mook H A, Chen G F, Luo J L, Wang N L and Dai P 2008 *Nature* **453** 899

[17] Nomura T, Kim S-W, Kamihara Y, Hirano M, Sushko P V, Kato K, Takata M, Shluger A L and Hosono H 2008 *Supercond. Sci. Technol.* **21** 125028

[18] Margadonna S, Takabayashi Y, McDonald M T, Brunelli M, Wu G, Liu R H, Chen X H, Kosmas Prassides 2008 arXiv/0806.3962

[19] Hosono H 2008 *J. Phys. Soc. Jpn.* **77**, *suppl*. C 1
[20] Ren Z-A, Che G-C, Dong X-L, Yang J, Lu W, Yi W, Shen X-L, Li Z-C, Sun L-L, Zhou F and Zhao Z-X 2008 *Europhys. Lett.* **83** 17002
[21] Kito H, Eisaki H and Iyo A 2008 *J. Phys. Soc. Jpn.* **77** 063707
[22] Sefat A S, Huq A, McGuire M A, Jin R, Sales B and Mandrus D 2008 *Phys. Rev. B* **78** 104505
[23] Cao C-G, Wang C, Zhu Z-W, Jiang S, Luo Y-K, Chi S, Ren Z, Tao Q, Wang Y and Xu Z 2008 arXiv/0807.1304
[24] Li Y K, Lin X, Zhu Z W, Chen H, Wang C, Li J L, Luo Y K, He M, Tao Q, Li H Y, Cao G H and Xu Z A 2008 arXiv/0808.3254
[25] Sefat A S, Jin R, McGuire M A, Sales B C, Singh D J and Mandrus D 2008 *Phys. Rev. Lett.* **101** 117004
[26] Leithe-Jasper A, Schnelle W, Geibel C and Rosner H 2008 arXiv/0807.2223
[27] Matsuishi S, Inoue Y, Nomura T, Yanagi H, Hirano M and Hosono H 2008 *J. Am. Chem. Soc.* **130** 14428
[28] Matsuishi S, Inoue Y, Nomura T, Hirano M, Hosono H 2008 *J. Phys. Soc. Jpn.* **77** 113709
[29] Han F, Zhu X, Mu G, Cheng P, Wen H-H 2008 *Phys. Rev. B* **78** 180503
[30] Tegel M, Johansson S, Weiss V, Schellenberg I, Hermes W, Poettgen R, Johrendt D 2008 arXiv/0807.1304
[31] Zhu X, Han F, Cheng P, Mu G, Shen B, Wen H-H 2008 arXiv/0810.2531
[32] Wu G, Xie Y L, Chen H, Zhong M, Liu R H, Shi B C, Li Q J, Wang X F, Wu T, Yan Y J, Ying J J, Chen X H 2008 arXiv/0811.0761
[33] Tarascon J M, Greene L H, Barboux P, McKinnon W R and Hull G W 1987 *Phys. Rev. B.* **36** 8393
[34] TOPAS, Version 3; Bruker AXS: Karlsruhe Germany, 2005
[35] Yanagi H, Kawamura R, Kamiya T, Kamihara Y, Hirano M, Nakamura T, Osawa H, and Hosono H 2008 *Phys. Rev. B* **77**, 224431
[36] Hiramatsu H, Katase T, Kamiya T, Hirano M and Hosono H 2008 *Appl. Phys. Express* **1** 101702

Effect of 3d Transition Metal Doping on the Superconductivity in CaFeAsF

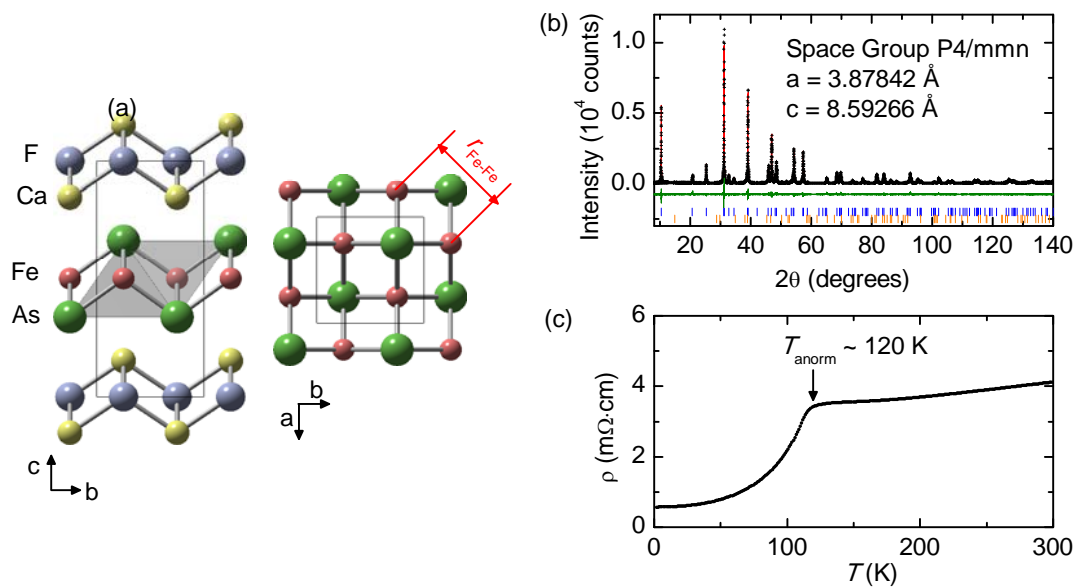

Figure 1 (a) Crystal Structure of CaFeAsF with ZrCuSiAs-type Structure (space group P4/nmm). (b) Powder XRD pattern of CaFeAsF (+): Red line indicates Rietveld fit pattern and Green line indicate different of observed and calculated pattern. Blue and red lines indicate diffraction positions for CaFeAsF and $Fe_2As$ impurity phases. (c) Temperature dependence of electrical resistivity of non-doped CaFeAsF.

Effect of 3d Transition Metal Doping on the Superconductivity in CaFeAsF

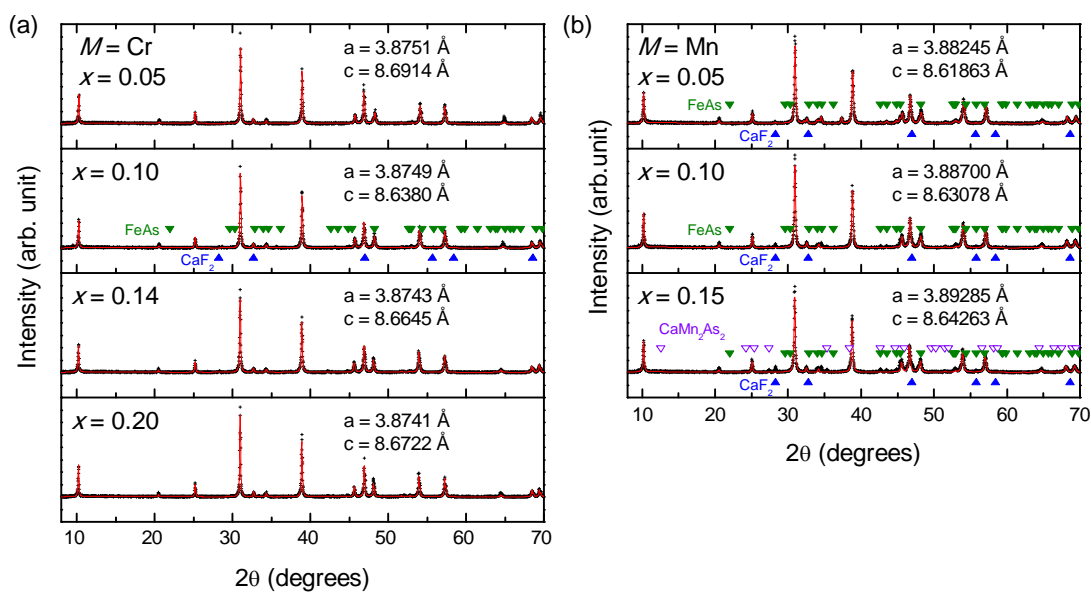

Figure 2 Powder XRD patterns of $CaFe_{1-x}Cr_xAsF$ with nominal $x = 0.05, 0.10, 0.14$ and $0.20$ (a) and $CaFe_{1-x}Mn_xAsF$ with nominal $x = 0.05, 0.10$ and $0.15$ (b): Crosses (+) denote observed patterns and red lines indicate calculated patterns for $CaFe_{1-x}M_xAsF$. Triangles are due to impurity phases ($CaF_2$, FeAs and $CaMn_2As_2$).

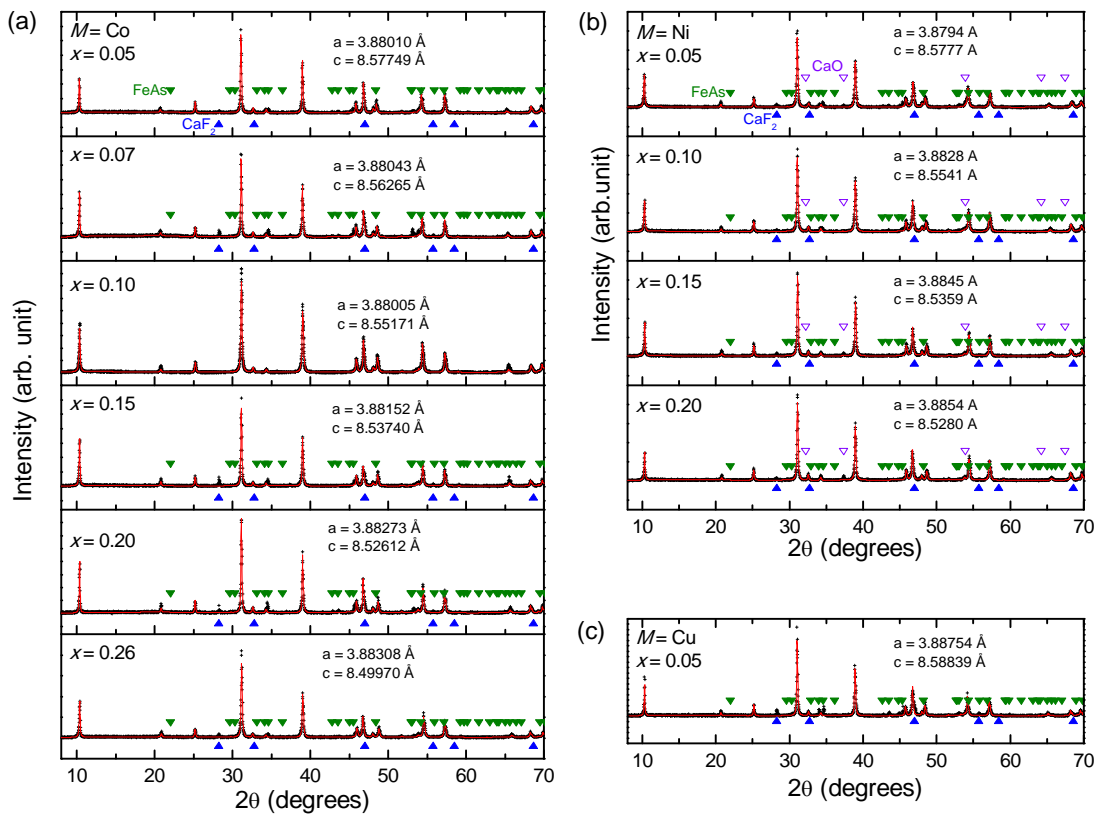

Figure 3 Powder XRD patterns of $CaFe_{1-x}Co_xAsF$ with nominal $x$ = 0.05, 0.07, 0.10, 0.15, 0.20 and 0.26 (a), $CaFe_{1-x}Ni_xAsF$ with nominal $x$ = 0.05, 0.10, 0.15 and 0.20 (b) and $CaFe_{0.95}Cu_{0.05}AsF$: Crosses (+) denote observed patterns and red lines indicate calculated patterns for $CaFe_{1-x}M_xAsF$. Triangles indicate diffraction positions of impurity phases ($CaF_2$, FeAs and CaO).

Effect of 3d Transition Metal Doping on the Superconductivity in CaFeAsF

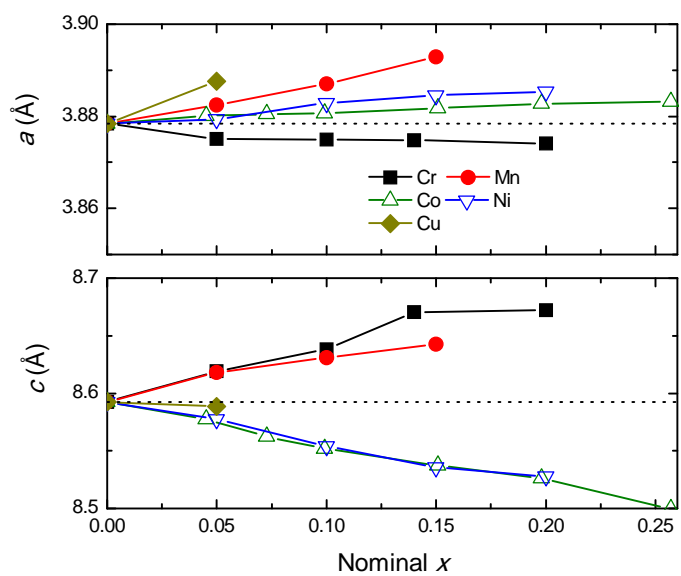

Figure 4 Change in lattice constants in CaFe$_{1-x}$M$_x$AsF with 3d transition metal ($M$ = Cr, Mn, Co, Ni and Cu) substitution.

Effect of 3d Transition Metal Doping on the Superconductivity in CaFeAsF

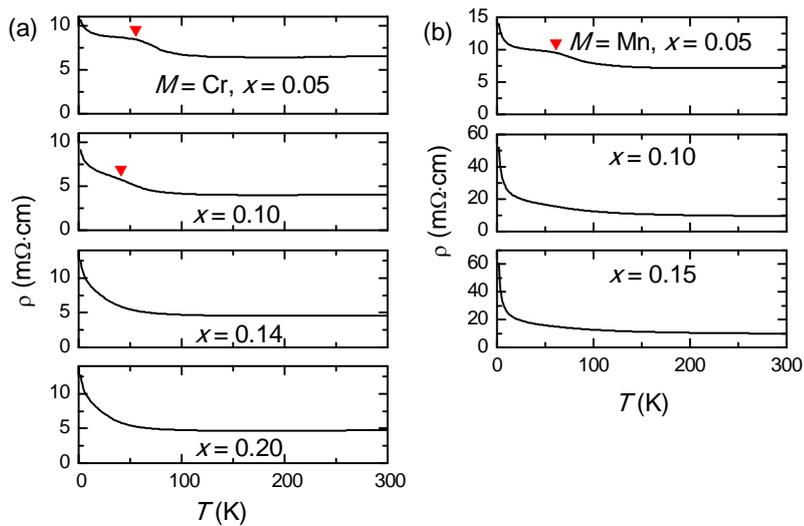

Figure 5 Temperature Dependence of electrical resistivity in CaFe$_{1-x}$TM$_x$AsF with 3d transition metal (TM = Cr, Mn) substitution. (a) CaFe$_{1-x}$Cr$_x$AsF (nominal $x$ = 0.05, 0.10, 0.14 and 0.20). (b) CaFe$_{1-x}$Mn$_x$AsF (nominal $x$ = 0.05, 0.10 and 0.15).

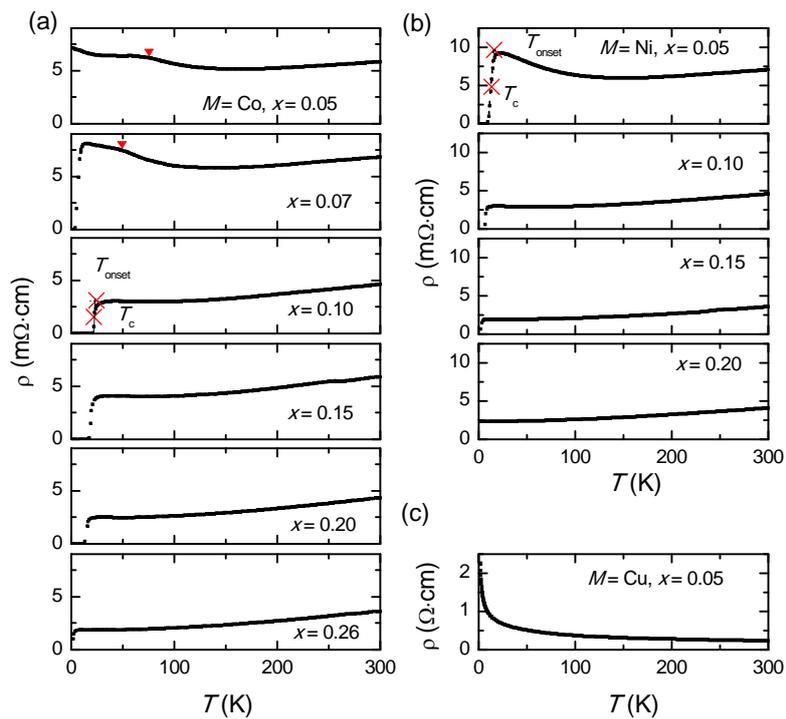

Figure 6 Temperature dependence of electrical resistivity in CaFe$_{1-x}$TM$_x$AsF with 3d transition metal (TM = Co, Ni, Cu) substitution (a) CaFe$_{1-x}$Co$_x$AsF (nominal $x$ = 0.05, 0.07, 0.10, 0.15, 0.20 and 0.26). (b) CaFe$_{1-x}$Ni$_x$AsF (nominal $x$ = 0.05, 0.10, 0.15 and 0.20). (c) CaFe$_{0.95}$Cu$_{0.05}$AsF.

Effect of 3d Transition Metal Doping on the Superconductivity in CaFeAsF

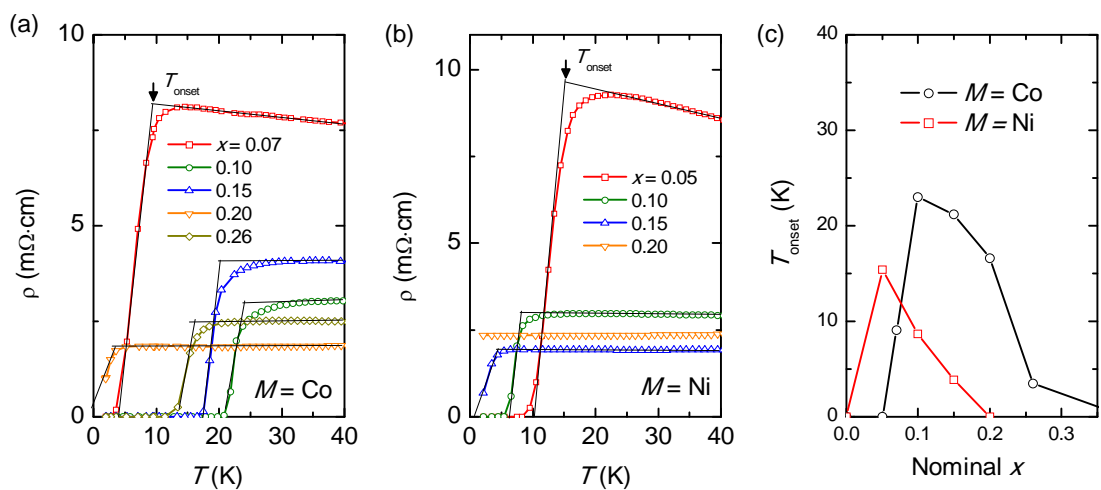

Figure 7 Magnified ρ-$T$ curves of CaFe$_{1-x}$Co$_x$AsF (a) and CaFe$_{1-x}$Ni$_x$AsF around $T_{onset}$ (b). (c) $T_{onset}$ as function of nominal $x$.